\documentclass[aps,prl,groupedaddress,reprint,showpacs,dvipdfmx]{revtex4-1}
\usepackage{graphicx}
\usepackage{tabularx}
\usepackage{dcolumn}
\usepackage{bm}
\usepackage{amsmath}
\usepackage{amssymb}
\usepackage{txfonts}
\usepackage{url}
\usepackage[usenames,dvipsnames]{color}
\definecolor{Gray}{gray}{0.0}
\definecolor{lightGray}{gray}{0.35}

\begin{document}
\title{
  A new {\it ab initio} modeling scheme for ion self-diffusion coefficient\\
  applied for $\varepsilon$-Cu$_3$Sn phase of Cu-Sn alloy
}
\author{Tom Ichibha$^{1,*}$}
\author{Genki Prayogo$^{2}$}
\author{Kenta Hongo$^{3,4,5,6}$}
\author{Ryo Maezono$^{1,6}$}
\affiliation{$^{1}$
  School of Information Science, JAIST, Asahidai 1-1, Nomi, Ishikawa,
  923-1292, Japan
}
\affiliation{$^{2}$
  School of Materials Science, JAIST, Asahidai 1-1, Nomi, Ishikawa,
  923-1292, Japan
}
\affiliation{$^{3}$
  Research Center for Advanced Computing Infrastructure, JAIST, Asahidai 1-1, Nomi, Ishikawa 923-1292, Japan
}
\affiliation{$^{4}$
  Center for Materials Research by Information Integration, Research and Services Division of Materials Data and Integrated System, National Institute for Materials Science, Tsukuba 305-0047, Japan
}
\affiliation{$^{5}$
  PRESTO, Japan Science and Technology Agency, 4-1-8 Honcho, Kawaguchi-shi, Saitama 322-0012, Japan
}
\affiliation{$^{6}$
  Computational Engineering Applications Unit, RIKEN, 
  2-1 Hirosawa, Wako, Saitama 351-0198, Japan
}
\affiliation{$^{*}$
  ichibha@icloud.com
}
\date{\today}
\begin{abstract}
  We present a new modeling scheme for ion self-diffusion coefficient,
  which broadens the applicable scope of {\it ab initio} approach.
  The essential concepts of the scheme are `domain division' and `coarse graining'
  of the diffusion network based on the barrier energies predicted by
  the {\it ab initio} calculation. The scheme was applied to evaluate Cu ion
  self-diffusion coefficient in $\varepsilon$-Cu$_3$Sn phase of Cu-Sn alloy,
  which is a typical system having long-range periodicity.
  The model constructed with the scheme successfully reproduces the experimental
  values in a wide temperature range.
\end{abstract}
\maketitle
\section{Introduction}
\label{sec.introduction}
Ion diffusion attracts the broad interests of the material researchers
because it dominates the various important phenomenas in the solid:~
corrosion, monotectoid, fracture, degradation and so on.
In order to elucidate the microscopic mechanism of the ion diffusion,
\textit{ab initio} calculation is one of the most powerful tools.
For example, it can evaluate the barrier energy of the diffusion route
with such techniques as nudged elastic band~(NEB) method\cite{2000HEN}.
Even the ion self-diffusion coefficient
\footnote{
  When the target ions are distributed ununiformly in the crystal,
  the distribution changes macroscopically with the thermal diffusion.
  This is called self-diffusion and the measure of the speed is
  self-diffusion coefficient. 
}
has been reported theoretically,
\cite{2012QIO,2009MANa,2009MANb,2010HUA,2011CHO,2014ZHAa}
combining the \textit{ab initio} predictions with modeling schemes
such as five-frequency model.\cite{1956LEC,1983KOIa,1983KOIb}
These theoretical works however target just simplest crystals mainly
due to the following two difficulties.

\vspace{2mm}
First, the system appearing in the practical situation
often requires a large size of unitcell because of
long-range periodicity, which makes the calculation expensive,
otherwise infeasible.
Such a problem is especially severe, for example,
when optimizing the ion path through the diffusion route
using NEB method because it requires a lot of force-field
calculations for a number of structures.
Secondly, it is very hard to count up and evaluate all of
the ion diffusion processes in most cases so the application range of
{\it ab initio} approach is limited within just simplest systems.

\vspace{2mm}
To overcome the difficulties, we introduced a couple of novel concepts,
which significantly simplifies the diffusion network based on the barrier energies.
First, suppose to classify the diffusion routes according to the barrier energies
into the three groups $I$-$III$ as the energies ordering like
$\Delta E_I\ll\Delta E_{II}\ll\Delta E_{III}$.
Since the barrier energy exponentially contributes the diffusion coefficient
({\it i.e.} Boltzmann factor), the diffusion routes (III) can be excluded
from the diffusion network. Then, the network may be divided into several
disjunct domains and the expensive calculation for the large unitcell
can be replaced by the several cheaper calculations for the small domains.
This is a concept to solve the first difficulty.
In addition, the diffusion network in each of the domains may be further
simplified by coarse-graining:
The vacancy can move almost freely through the diffusion routes (I)
compared with moving through the routes (II), so the ion sites
connected by the routes (I) can be represented by just a site.
Eventually, the diffusion network falls into coarse-grained one
with the representative sites, where it is much easier to
count up the diffusion processes. 

\vspace{2mm}
We established a modeling scheme based on the above concepts
for the example of the Cu ion self-diffusion coefficient
in $\varepsilon$-Cu$_3$Sn phase of Cu-Sn alloy.
This is a typical system having long-range periodicity
and there is abundant experimental data for the Cu self-diffusion coefficient.
\cite{2006CHA,2009CHA,2013YAN,2013WAN,2011KUM,2011PAU}
From simulation side, classical molecular dynamics (MD) was applied\cite{2012GAO}
but it overestimated the experimental values by a digit.
On the other hand, the model constructed with our scheme successfully
reproduced them in a wide temperature range.

\vspace{2mm}
This paper consists of the following sections:  
In the next section (Formalism), we introduce
the equations for the physical quantities
relevant to the ion diffusion and also define 
the technical terms and the notation rules.
In the 3rd section (Domain division),
we explain how to divide the diffusion network
into the multiple types of the simple domains,
based on the barrier energies.
In the 4th section (Details and results of {\it ab initio} calculation),
we present the predictions for the quantities
(ex. barrier energy, vacancy formation energy)
calculated with the density functional calculation
The details of the calculations are also given here.
In the 5th section (Coarse graining),
we discuss how to represent
the multiple ion sites with just a site
and coarse-grain the diffusion network,
based on the barrier energies.
In the 6th section (Formula of self-diffusion coefficient),
We discuss how to model the self-diffusion coefficient  
when there are two or more types of diffusion routes.
In the 7th section (Evaluation of correlation factor),
it is deeply discussed how to count up the possible
diffusion processes in the simplified diffusion network
and evaluate the correlation factor with some approximations.
In the 8th section (Results and discussions),
the validity of the model constructed with our
scheme is discussed, compared with the experimental values
for the self-diffusion coefficient or the correlation factor
for ideal hexagonal 2-dimensional lattice.
In the 9th section (Summary and prospects),
we summarize this paper and refer to the advantage
of our scheme from {\it ab initio} MD.

\section{Formalism}
\label{sec.theory}
The self-diffusion coefficient $D(w)$ represents the macroscopic ion flow
in a direction $w$ driven by self-diffusion. $D(w)$ is obtained by evaluating
the ion jump, which is a minimum process of the ion diffusion.
We index the diffusion routes connecting to a site with $j$
and define $\theta_j(w)$ as the angle of the ion jump thorough route $j$
from direction $w$.
We also define  $D^*_j$ as the angle-independent contribution of the ion jump
to the self-diffusion coefficient (explained in the next paragraph)
and the self-diffusion coefficient $D(w)$ is evaluated from the following
equation:\cite{2007MEH}
\begin{eqnarray}
  D\left( w \right) = \sum\limits_j {\frac{1}{2}{D^*_j}
    {{\cos }^2}{\theta _j}\left( w \right)}.
  \label{eq.self}
\end{eqnarray}

\vspace{2mm}
$D^*_j$ consists of the four quantities appeared
in the following equation (index $j$ is omitted):~\cite{2007MEH}
\begin{eqnarray}
  {D^*} = d^2\cdot f\cdot C_v\cdot\eta.
  \label{eq.diffusion1}
\end{eqnarray}
$d$ is the distance of the diffusion route.
$C_v$ is the vacancy formation rate given as the Boltzmann
factor of the vacancy formation energy $\Delta E_{\rm vac}$.
$\eta$ is the jump rate depending on the barrier
energy $\Delta E_{\rm barrier}$ and the vibration frequency
$\nu$ of the attentional ion (tracer) in the jump direction:~
\begin{eqnarray}
  \eta = \nu\cdot\exp\left({-\Delta E_{\rm barrier}/{k_B}T}\right).
\end{eqnarray}
$f$ is called correlation factor and explained in the next paragraph.
The evaluation of the vacancy formation energy $\Delta E_{\rm vac}$
needs the chemical potential $\mu_{\mathrm{Cu}}$ of Cu ion.
We calculated it for face-centered Cu mono-crystal and
got the $\Delta E_{\rm vac}$ from the following equation:~
\begin{eqnarray}
  \Delta {E_{\rm vac}} &=& {E_{\rm vac}} - {E_{\rm perfect}} + {\mu _{\rm Cu}}.
  \label{eq.vacancy}
\end{eqnarray}
Here, $E_{\rm perfect}$ is the total energy of perfect crystal
and $E_{\rm vac}$ is that of the crystal including a vacancy defect.

\vspace{2mm}
The evaluation of the correlation factor $f$ is
the most awful part for the modeling of self-diffusion coefficient.
The factor reflects the position exchanges sequentially
occurring between the vacancy and the tracer:
After the tracer jumping via the vacancy, it is at the behind of the tracer.
Thus, the vacancy tends to pull back the tracer at the next ion jump
around the reverse direction of the ion jump. 
In addition, such a pull-back can
occur again and again after the 1st pull-back,
always around the reverse direction of the preceding tracer move.
The correlation factor $f$ represents how much the sequential pull-backs
affect the self-diffusion coefficient in total.
Since the 1st pull-back denying the tracer jump has
the largest contribution to the correlation factor,
it always reduces $D^*_j$ ({\it i.e.} $0 \le f \le 1$).

\vspace{2mm}
To evaluate correlation factor, we have to count up
all of the vacancy tracks from the one to the next
position exchange between the tracer and the vacancy.
This is the identity of the first difficulty.
The contribution from each of the pull-backs to
the self-diffusion coefficient is given as average cosine:
\begin{eqnarray}
  \left\langle {\cos \theta } \right\rangle ^{\left( n \right)}
    = \sum\limits_k^Z {{P_k}\cos {\theta _k}}.
    \label{eq.average_cosine}
\end{eqnarray} 
Here, $P_k$ is the sum of the realization probability
of the vacancy tracks causing the pull-back from site $k$. 
$\theta_k$ is the angle of the tracer move at the pull-back
from the preceding tracer move. We also define
$n$th-order average cosine
${\left\langle {\cos \theta } \right\rangle ^{\left( n \right)}}$,
which represents how much the first ion jump is denied by
the $n$ times pull-backs in average. The correlation factor
$f$ consists of them as the following equation:~ 
\begin{eqnarray}
  f = 1 + 2\sum\limits_{n = 1}^\infty  {{{\left\langle {\cos \theta }
        \right\rangle }^{\left( n \right)}}}.
  \label{eq.correlation_factor}
\end{eqnarray}
`$(n+1)$ times of pull-back' is interpreted as one extra pull-back
occurring after `$n$ times of pull-back' so
${\left\langle {\cos \theta } \right\rangle ^{\left( n \right)}}$
is given as
\begin{eqnarray}
  {\left\langle {\cos \theta } \right\rangle ^{\left( {n + 1} \right)}}
  = {\left\langle {\cos \theta } \right\rangle ^{\left( n \right)}}
  \left\langle {\cos \theta } \right\rangle.
  \label{eq.chain}
\end{eqnarray}
Especially when there in only one diffusion route,
\footnote{
  For example, if you consider the ion jumps between
  the sites having the shortest distance, there are
  only one type of diffusion routes. On the other hand,
  if you allow the ion jumps between the sites having
  the second shortest distance alos, there are two
  types of diffusion routes.
}
the following relationship is hold:\cite{2007MEH}
\begin{eqnarray}
  {\left\langle {\cos \theta } \right\rangle ^{\left( n \right)}} = {\left\langle {\cos \theta } \right\rangle ^n}.
    \label{eq.chain2}
\end{eqnarray}
This equation is generalized for multiple diffusion routes
later in section (formula of correlation factor).

\section{domain division}
\label{systems}
\begin{figure}[htbp]
  \centering
  \includegraphics[width=1.0\hsize]{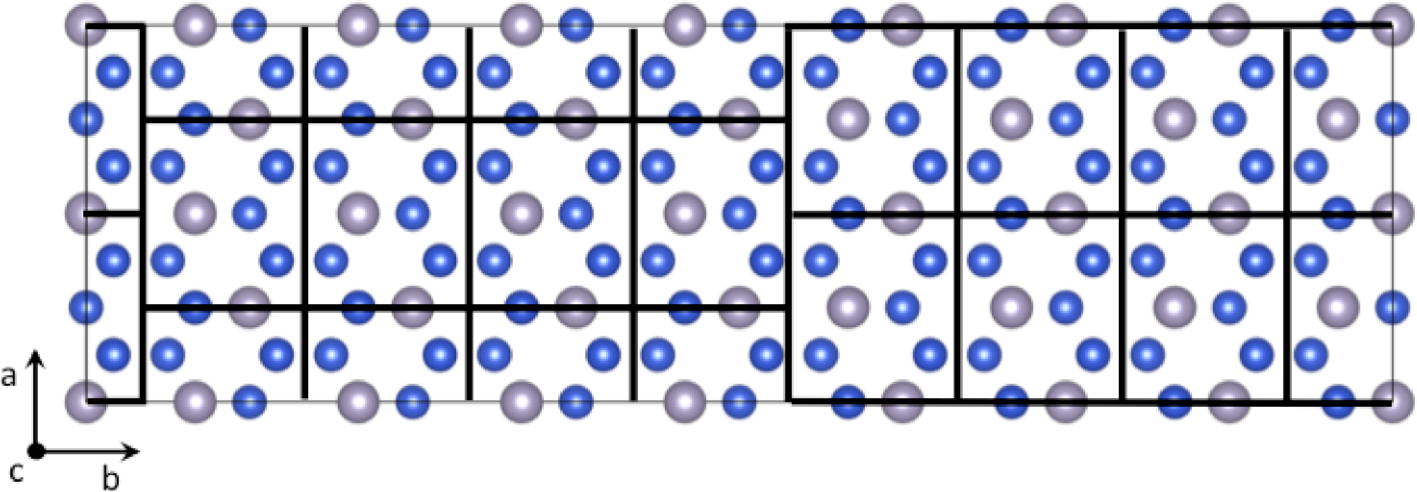}
  (a)
  \includegraphics[width=1.0\hsize]{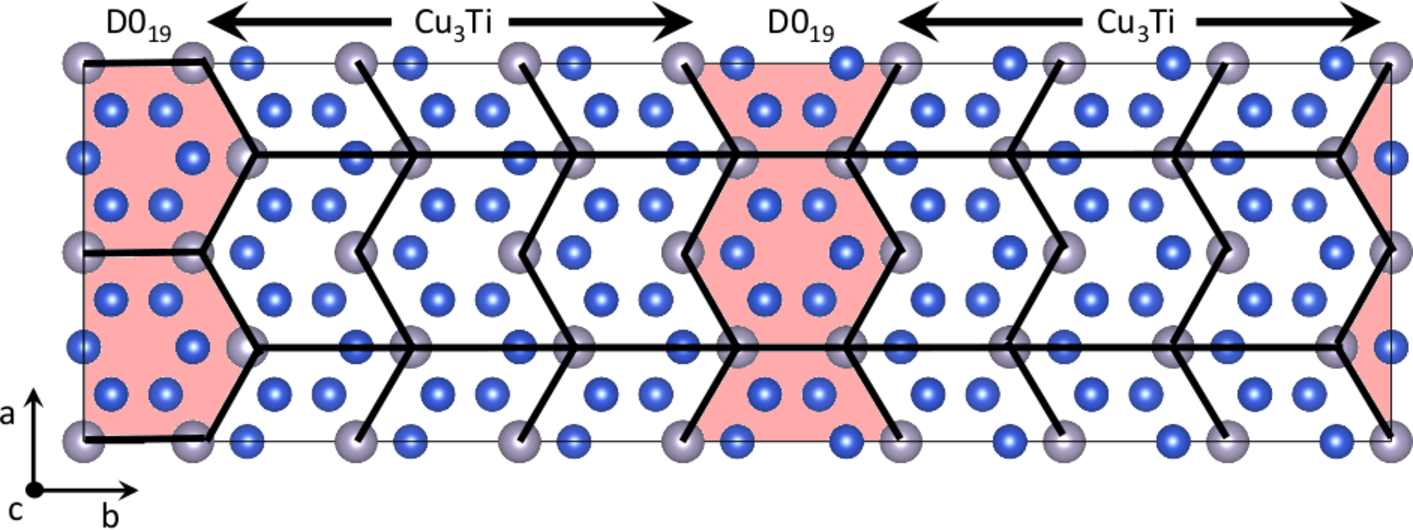}
  (b)
  \caption{
    The crystal structure of $\varepsilon$-Cu$_3$Sn\cite{1978SAN}
    The blue balls represent Cu ions and the gray balls do Sn ions.
    Figure (a) shows the conventional interpretation for
    the long-range periodic structure. Cu$_3$Ti-type unitcell
    (surrounded by thick line) is shifted
    by 1/2 period along $a$ direction every $M/2$
    ($M$:~periodicity) unitcells lining up along $b$ direction. 
    Figure (b) shows our new interpretation.
    $(M-2)/2$ unitcells of Cu$_3$Ti-type structure and
    one D0$_{19}$-type unitcell appear periodically along $b$ direction.
    The pictures are made with VESTA.\cite{2011MOM}
  }
  \label{fig.epsilon}
\end{figure}

$\varepsilon$-Cu$_3$Sn phase has, it is reported,
long-range periodic structure (Fig.~\ref{fig.epsilon}a).
It has been recognized that the structure consists of
the unitcells of Cu$_3$Ti-type structure with shifted by
1/2 in fractional coordinate along $a$-axis,
every $M/2$ unitcells lining up along $b$-axis~(Fig.~\ref{fig.epsilon}a).
The various periodicities $M$ are reported experimentally
as the even numbers from 2 to 12.
\cite{1978SAN,2009SAN}
Based on the {\it ab initio} predictions,
it is newly found that the ion jump though the diffusion route
passing between the Sn sites hardly contributes to the self-diffusion
because their barrier energies are much higher than the others.
After excluding these routes from the diffusion network, 
it is divided into the two types of simple domains,
Cu$_3$Ti-type and D0$_{19}$-type, as shown in Fig.~\ref{fig.epsilon}b.
The both domains exist in the ratio of ($M$-2):2 so
the self-diffusion coefficient of the target phase is given as
the weighted average of them with the existing ratio:
\begin{eqnarray}
   {D_{\varepsilon  - {\rm{C}}{{\rm{u}}_{\rm{3}}\rm{Sn}}}} 
   = \frac{{M - 2}}{M}{D_{{\rm{C}}{{\rm{u}}_{\rm{3}}}{\rm{Ti}}}}
   + \frac{2}{M}{D_{{\rm{D0_{19}}}}}.
   \label{totalD}
   \label{eq.M}
\end{eqnarray}

\section{details and results of {\it ab initio} calculation}
We employed 2$\times$2$\times$2 supercell for
both Cu$_3$Ti-type and D0$_{19}$-type domains
to reduce the spurious interaction between the vacancies.
The lattice parameters are fixed at the experimental values.\cite{1978SAN}
We used VASP\cite{1996KRE} for the density functional calculation
to evaluate the physical quantities.
The barrier energies are evaluated with
climbing-nudged elastic band (c-NEB) method \cite{2000HEN}
implemented in VASP. \cite{1996KRE}
c-NEB needs the interpolated structures
between the both edges of the diffusion path
as inputs and optimizes them with connected by the virtual springs.
c-NEB guarantees that one of the structures is positioned
at the exact saddle point of the given potential surface,
in which c-NEB is superior to the original NEB method.\cite{2011JON}
We prepared 15 interpolated structures and set the spring
coefficient as 5~eV/$\AA^2$.
We used PBE functional \cite{1996PER} and PAW pseudopotentials
provided in VASP\cite{1999KRE} for all of the calculations.
We determined the cutoff energy and k-mesh for the total energy
to be converged within 0.5~kJ/unitcell for the both
Cu$_3$Ti-type and D0$_{19}$-type domains.
We calculated the vibrational frequency of the tracer in the jump
direction with the harmonic approximation of the potential surface.

\begin{figure}
  \includegraphics[width=\hsize]{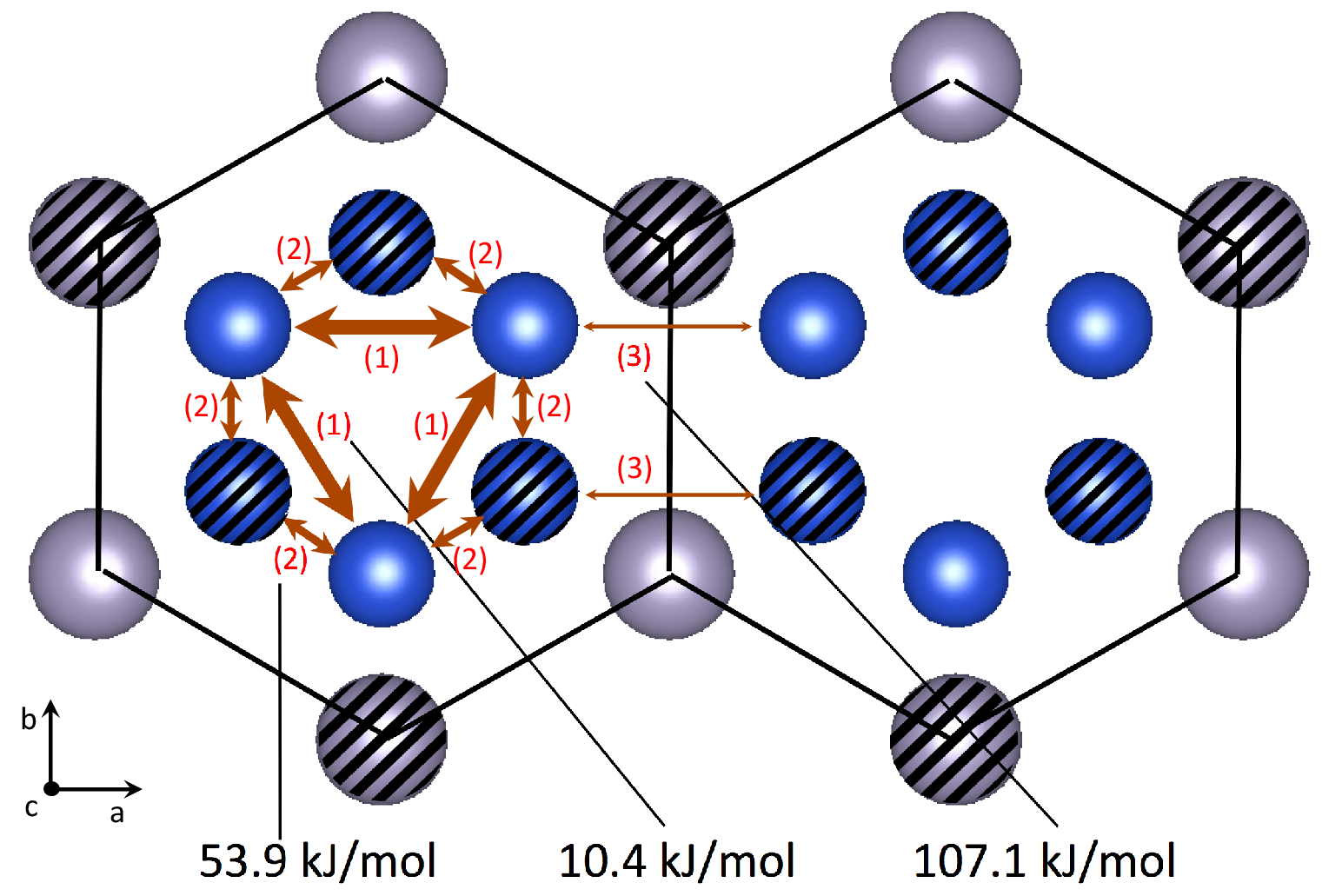}
  \caption{
    The diffusion routes and the barrier energies
    of Cu ion in D0$_{19}$-type domain.
    The diffusion routes are shown as the two-way
    arrows with the values of the barrier energies.
    The blue balls represent Cu ions and the gray balls
    do Sn ions. The hatched balls are located on c=1/2 plane
    and the unhatched ones are on c=0,~1 plane in fractional coordinate.
    The ion diffusion occurs in the 1-dimensional tubes partitioned by
    the thick black lines, since the energy barriers of the route 3
    across the lines are much higher than the others.
    This picture is made with VESTA\cite{2011MOM}.
  }
  \label{fig.d019}
\end{figure}

\begin{figure}
  \includegraphics[width=\hsize]{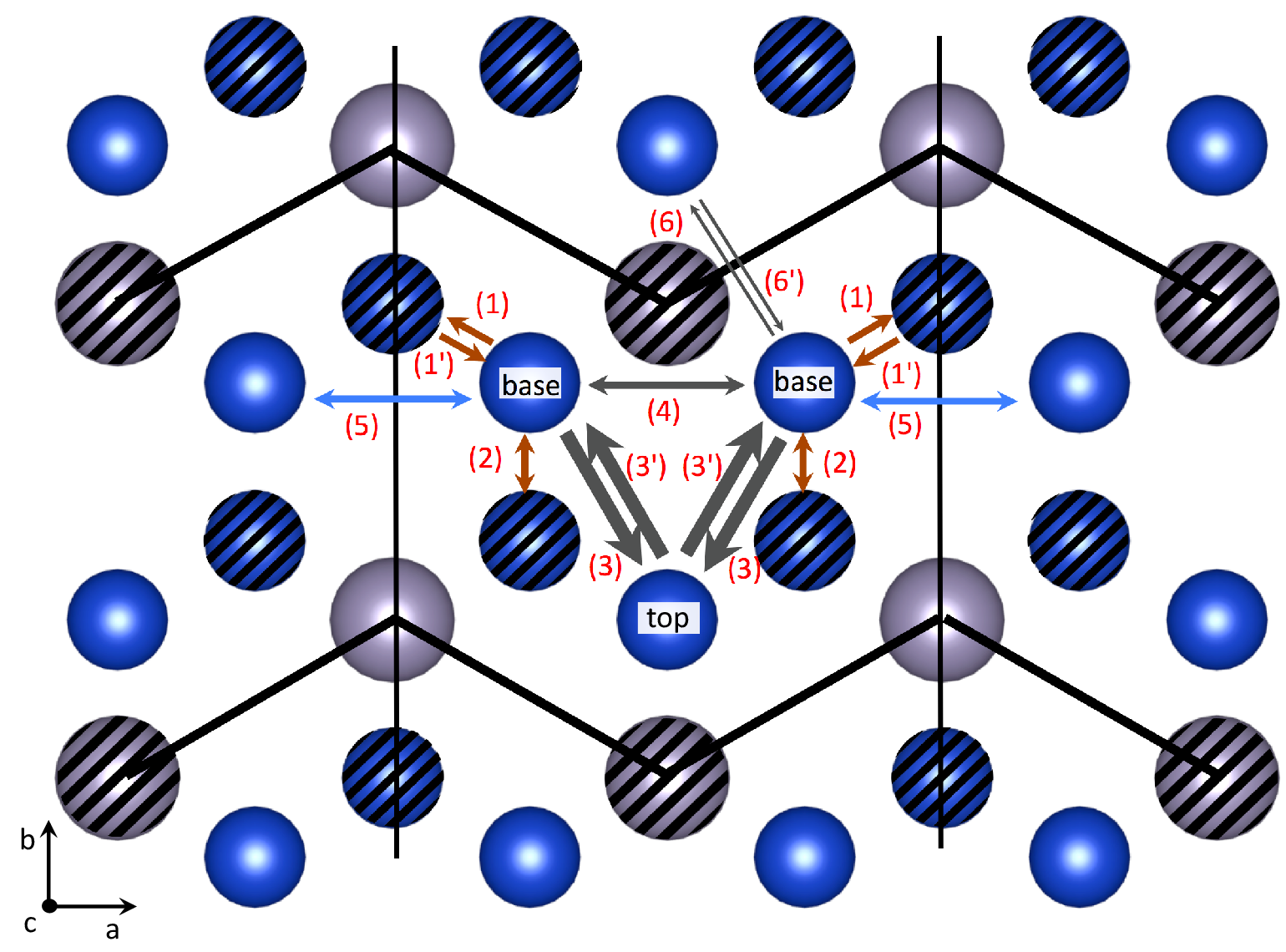}
  \caption{
    The diffusion routes of Cu ion in Cu$_3$Ti-type domain.
    The blue balls represent Cu ions and gray balls do Sn ions.
    The hatched balls are located on $c$=1/2 plane and
    the unhatched ones are on $c$=0,~1 plane in fractional coordinate.
    The diffusion routes are shown as the red arrows (1,1,',2) and
    blue ones (5) and they are named route $\tilde c$ and $a$ respectively.
    The both-side arrow represents that the ion jumps in normal and reverse
    directions are equivalent.
    The ion diffusion occurs in the 2-dimensional layers partitioned by
    the thick black lines, since the energy barriers of the route 6,6'
    across the lines are much higher than the others.
    This picture is drawn with VESTA\cite{2011MOM}.
  }
  \label{fig.cu3ti}
\end{figure}

\vspace{2mm}
The diffusion routes in D0$_{19}$-type (Cu$_3$Ti-type) domain
is shown in Fig.~\ref{fig.d019} (Fig.~\ref{fig.cu3ti}).
The diffusion barriers for D0$_{19}$-type domain are written in the figure.
Those for the Cu$_3$Ti-type domain are written in Tab.~\ref{tab.cu3ti}
with the vibrational frequencies $\nu$ in the jump direction and jump rate $\eta$
at 300~K and 423~K. It can be observed from the table that the route 3 of
D0$_{19}$-type domain and the route 6,~6' of Cu$_3$Ti-type have the highest
barrier energies, which is the ground of the domain division of the diffusion network.
D0$_{19}$-type domain has only one type of Cu site and
the vacancy formation energy was calculated as 16.60~kJ/mol.
Cu$_3$Ti-type domain has two types of Cu site denoted by
'top' and 'base' in Fig.~\ref{fig.cu3ti}.
The vacancy formation energies are calculated as 25.94~kJ/mol
for 'top' and 21.71~kJ/mol for 'base'.

  \begin{table*}[htbp]
    \caption{
      The list of the physical quantities relevant to
      the ion diffusion in Cu$_3$Ti-type domain.
      The indices $i$ correspond to the ones of the
      diffusion routes shown in Fig.~\ref{fig.cu3ti}.
      $\Delta E_i$ is the energy barrier,
      $\nu_i$ is the vibration frequency towards the jump direction,
      and $eta_i$ is the ion jump frequencies at 300~K and 423~K.
    }
    \label{tab.cu3ti}
  \begin{center}
    \begin{tabular}{ccccc}
      \hline
      Path $i$ & $\Delta E_i$~[kJ/mol] & $\nu_i$~[sec.$^{-1}$] & $\eta_i$~(300~K)~[sec.$^{-1}$] & $\eta_i$~(423~K)~[sec.$^{-1}$] \\
      \hline
      1 & 54.59 & 3.675$\times 10^{12}$ & 1.151$\times 10^{3}$ & 6.680$\times 10^{5}$ \\
      1'& 58.82 & 3.575$\times 10^{12}$ & 2.049$\times 10^{2}$ & 1.948$\times 10^{5}$ \\
      2 & 69.65 & 4.152$\times 10^{12}$ & 3.096$\times 10^{0}$ & 1.041$\times 10^{4}$ \\
      3 & 21.69 & 3.155$\times 10^{12}$ & 5.288$\times 10^{8}$ & 6.625$\times 10^{9}$ \\
      3'& 25.91 & 2.542$\times 10^{12}$ & 7.828$\times 10^{7}$ & 1.605$\times 10^{9}$ \\
      4 & 60.19 & 4.138$\times 10^{12}$ & 1.369$\times 10^{2}$ & 1.527$\times 10^{5}$ \\
      5 & 51.60 & 3.618$\times 10^{12}$ & 3.748$\times 10^{3}$ & 1.536$\times 10^{6}$ \\
      6 & 99.44 &                  -- &                 -- &                  -- \\
      6'&103.66 &                  -- &                 -- &                  -- \\
      \hline
    \end{tabular}
  \end{center}
\end{table*}

\section{Coarse graining}
\label{sec.modeling}
\begin{figure}
  \includegraphics[width=\hsize]{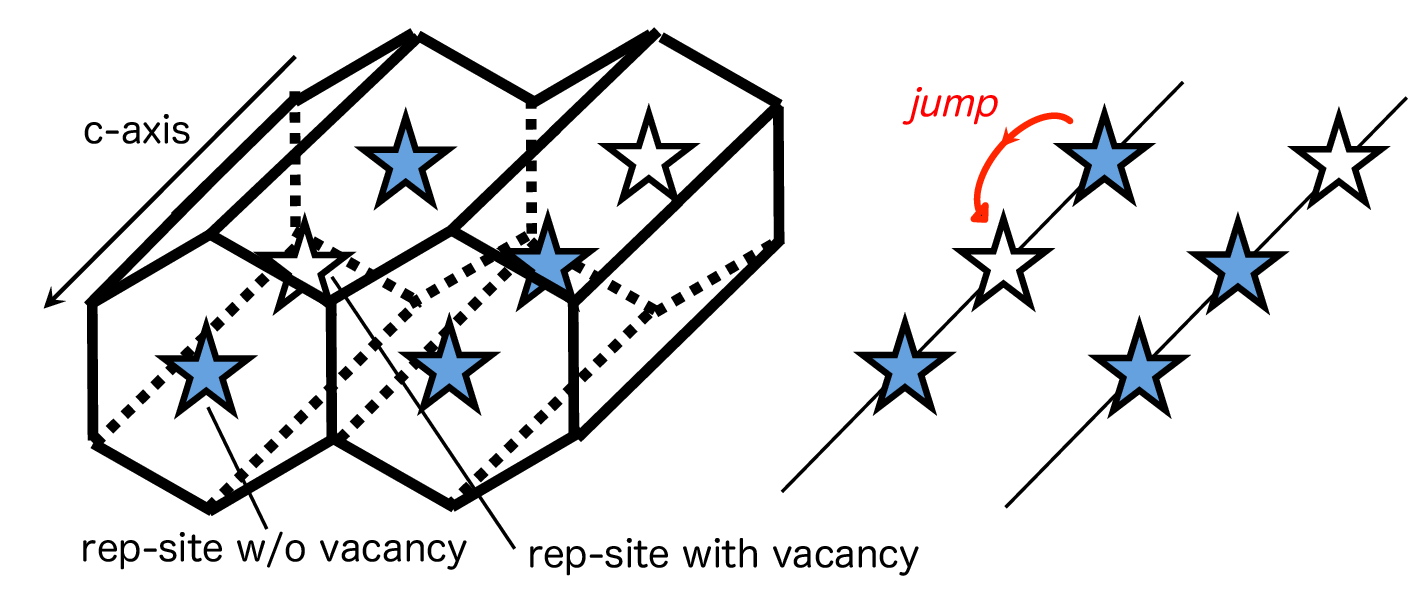}
  \caption{
    The left figure shows the coarse-grained diffusion network
    of D0$_{19}$-type domains with the rep-sites.
    The star represents rep-site and the open one includes the vacancy.
    Since the tracer seldomly jumps perpendicular to $c$-axis,
    the diffusion network is almost 1-dimensional.
  }
  \label{fig.d019_model}
\end{figure}

In the structure of $D0_{19}$-type domain,
the hexagonal tubes consisting of Cu sites
surrounded in the lines connecting Sn sites.
Each of the tubes consists of the laminated
Cu ions arranged in the triangle shape
located in the different $ab$-planes
by 1/2 in fractional coordination (hatched and unhatched sites).
Since the barrier energy of route 3 connecting the two tubes
is much higher than the others, the ions diffuse in each of
the independent tubes.
In addition, route 1 has much lower barrier energy than
route 2 so the vacancy can move inside of the triangle
through route 1 much more easily than moving across
the triangle through route 2.
Therefore, it would be possible to represent the triangle
with just a site, and the diffusion network is coarse-grained
into the 1-dimensional one long $c$-axis
composed of the representative sites, named rep-site
(a schematic picture given in Fig.~\ref{fig.d019_model}).
The correlation factor of 1-dimensional diffusion
is known to be zero in the textbook\cite{2007MEH} and hence
\begin{eqnarray}
  D_{\rm{D0_{19}}}=0.
  \label{eq.dd0190}
\end{eqnarray}

\begin{figure}
  \includegraphics[width=\hsize]{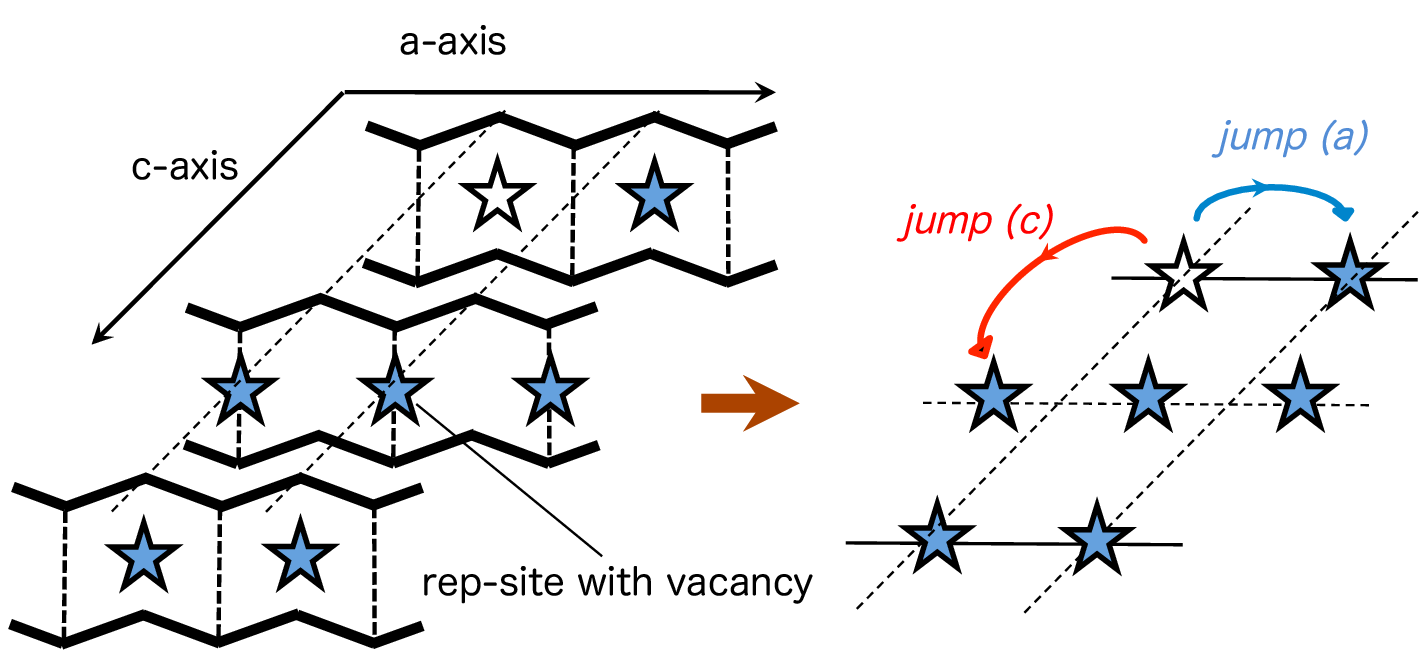}
  \caption{
    The left figure shows the coarse-grained diffusion
    network of Cu$_{3}$Ti-type domain with the rep-sites.
    The star represents rep-site and the open one includes the vacancy.
    Since the tracer seldomly jumps perpendicular to $ac$-plane,
    the diffusion network is almost 2-dimensional.
    The rep-sites are connected with two types of routes
    name route $a$ and $\tilde c$. Route $a$ is parallel to
    $a$-axis and route $\tilde c$ deviates from $c$-axis.
  }
  \label{fig.cu3ti_model}
\end{figure}

\vspace{2mm}
The diffusion network of Cu$_3$Ti-type domain can be also coarse-grained
with the same analogy yet into the 2-dimensional one:
The barrier energy of route 6,6’ is much higher than the others
so the ion diffusion occurs in the layers partitioned by the thick
lines written in Fig.~\ref{fig.cu3ti}.
In addition, the barrier energies of route 3,3’ are
much lower than the others, so the three sites connected
by route 3,3’ can be represented by a rep-site.
The vacancy formation rate of the rep-site would reasonably
be evaluated as
\begin{eqnarray}
  C_{\rm{v}}^{\left( {{\rm{rep}}} \right)}
  = C_{\rm{v}}^{\left( {{\rm{top}}} \right)}
  + 2C_{\rm{v}}^{\left( {{\rm{base}}} \right)} .
\end{eqnarray}
After the coarse-graining with using a rep-site,
the diffusion network becomes the 2-dimensional
one consisting of two types of diffusion
routes $a$,~$\tilde c$ shown in Fig.~\ref{fig.cu3ti_model}.
(Noted that route $a$ is parallel to $a$-axis but route
$\tilde c$ deviates from $c$-axis.)
Route $a$ consists of route 5 and route $\tilde c$ consists of
route 1,1',2 so the jump rates through routes $a$,~$\tilde c$
can be given as:
\begin{eqnarray}
  {C_v^{\mathrm{(rep)}}}\cdot{\eta _a}
  &=& {C_v^{\mathrm{(base)}}}\cdot{\eta _5} ,\label{eq.eta_t1} \\
  {C_v^{\mathrm{(rep)}}}\cdot{\eta _{\tilde c}} &=& {C_v^{\mathrm{(top)}}}\cdot{\eta _1} 
  + {C_v^{\mathrm{(base)}}}\cdot\left( {{\eta _{1'}} + {\eta _2}} \right). \label{eq.eta_s1} 
\end{eqnarray}
We define ${\gamma ^{\left( {{\rm{top}}} \right)}}$,
${\gamma ^{\left( {{\rm{top}}} \right)}}$ as:
\begin{eqnarray}
  \gamma^{\mathrm{(top)}}\equiv C_v^{\mathrm{(top)}}/C_v^{\mathrm{(rep)}},
  \;\;\gamma^{\mathrm{(base)}}\equiv C_v^{\mathrm{(base)}}/C_v^{\mathrm{(rep)}}.
\end{eqnarray}
and substitute
${\gamma ^{\left( {{\rm{top}}} \right)}}$,
${\gamma ^{\left( {{\rm{base}}} \right)}}$
into eq.~\eqref{eq.eta_t1},\eqref{eq.eta_s1}.
Then, we can get the effective jump rates $\eta_a$,~$\eta_{\tilde c}$ as:
\begin{eqnarray}
  {\eta _a} = \gamma^{\mathrm{(base)}} {\eta _5},\;\;
  {\eta _{\tilde c}} = \gamma^{\mathrm{(top)}} {\eta _1}
  + \gamma^{\mathrm{(base)}} \left( {{\eta _{1'}} + {\eta _2}} \right).
\end{eqnarray}
Noted that ${\gamma ^{\left( {{\rm{top}}} \right)}}$ and
${\gamma ^{\left( {{\rm{base}}} \right)}}$ correspond to
the probabilities of that the vacancy found in a top
(base) site after the vacancy wondering in a rep-site
for a sufficiently long time.

\vspace{2mm}
Denoting the self-diffusion coefficients in direction $a,~c$ as $D(a),~D(c)$,
the angle averaged self-diffusion coefficient of $\varepsilon$-Cu$_3$Ti-type
domain can be evaluated as
\begin{eqnarray}
  {D_\mathrm{C{u_3}Ti}} = 1/3 \cdot \left( {{D(a)} + {D(c)}} \right)  .
 \label{cu3ti_diff}
\end{eqnarray}
Once the $D^*_a$ and $D^*_{\tilde c}$ are obtained
from eq.~\eqref{eq.diffusion1}, they are projected
into $D(a)$ and $D(\tilde c)$ based on eq.~\eqref{eq.self}:
\begin{eqnarray}
  {D(a)} &=& D_a^* + 2D_{\tilde c}^*{\cos ^2}{\theta _{a\tilde c}}, \\
  {D(c)} &=& 2D_{\tilde c}^*{\cos ^2}{{\theta _{c \tilde c}}}.
  \label{each_axis}
\end{eqnarray}
Here, ${\theta_{a\tilde c}}$~(${\theta_{c \tilde c}}$) is the angle
between the direction $a$~($c$)-axis and route $\tilde c$.

\section{formula of correlation factor}
\begin{figure}
  \includegraphics[width=\hsize]{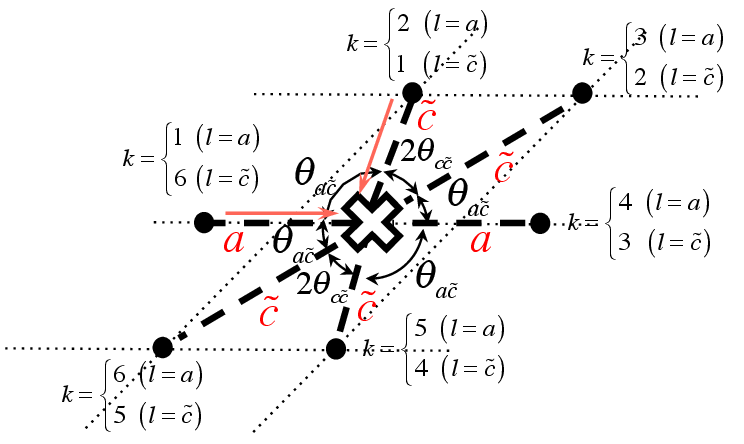}
  \caption{
    The tracer just moved to the site denoted
    by open cross through route $a$ or $\tilde c$.
    The surrounding 6 sites are indexed with $k=1\sim6$
    in a clockwise fashion beginning from
    the preceding tracer position before moving.
  }
  \label{fig.conv}
\end{figure}
In order to obtain $D_{a,\tilde c}^*$ from eq.~\eqref{eq.diffusion1},
we have to evaluate the correlation factors for both of the route $a$
and route $\tilde c$. We established the equations alternating
eq.~\eqref{eq.chain2} for multiple types of diffusion routes.

\vspace{2mm}
Before explaining about it, we will confirm the notation
rules used in this paper. Suppose that the tracer
(open cross) has just moved through route $l=a$
or $\tilde c$ shown as the arrows towards the tracer in Fig.~\ref{fig.conv}.
The six sites surrounding the tracer are indexed with $k=1-6$
in clockwise order beginning from the site where the tracer
was positioned before moving.
$\theta_k$ in eq.~\eqref{eq.average_cosine} is the angle
between the directions of the sequential tracer moves.
Hence, for example, $\theta_2={\theta _{a\tilde c}}$ for ion move $l=a$
and $\theta_2=2{\theta _{c\tilde c}}$ for ion move $l=\tilde c$.
For convenience, the $l$ dependency is noted as $\theta_k^{(l)}$
and a new function $L^{(l)}(k)$ is also introduced, which returns
the type of the diffusion route ($a$ or $\tilde c$)
connecting site $k$ and the tracer position.
The return values are listed specifically as:
$L^{(a)}(k=1,4)=a$, $L^{(a)}(k=2,3,5,6)=\tilde c$,
$L^{(\tilde c)}(k=3,6)=a$, and $L^{(a)}(k=1,2,4,5)=\tilde c$.

\vspace{2mm}
The $n$th-order average cosign
$\left\langle {\cos \theta } \right\rangle _{l = a,~\tilde c}^{\left( {n } \right)}$
may be built as the below recurrence formula similar to the idea for eq.~\eqref{eq.chain2}:
\begin{eqnarray}
  \left\langle {\cos \theta } \right\rangle _{l = a,~\tilde c}^{\left( {n + 1} \right)} =
  \sum\limits_k {\frac{{d_{L^{(l)}(k)}}}{{d_l}}
    P_k^{(l)}\cdot\cos {\theta_k^{(l)}}\cdot
    \left\langle {\cos \theta } \right\rangle _{{L^{(l)}(k)}}^{\left( n \right)}} .
  \label{eq.aveCos}
\end{eqnarray}
This equation can be solved analytically and, of course, numerically also.    
The correlation factors for the ion jumps
through route $a$ and route $\tilde c$ is given as same
as eq.~\eqref{eq.correlation_factor}
\begin{eqnarray}
  {f_{a,~\tilde c}} = 1 + 2\sum\limits_{n = 1}^{n_{\mathrm{max}}}
  {\left\langle {\cos \theta } \right\rangle _{a,~\tilde c}^{\left( n \right)}}
  \approx 1 + 2\sum\limits_{n = 1}^{\infty}  {\left\langle {\cos \theta }
    \right\rangle _{a,~\tilde c}^{\left( n \right)}},
  \label{eq.corrFac}
\end{eqnarray} 
which is same as \eqref{eq.correlation_factor}. Here, $n_{max}$ should be
large enough for $f_{a,~\tilde c}$ to be converged.

\section{Evaluation of correlation factor}
\label{sec.eval_corr}
\begin{figure}
  \centering
  \includegraphics[width=\hsize]{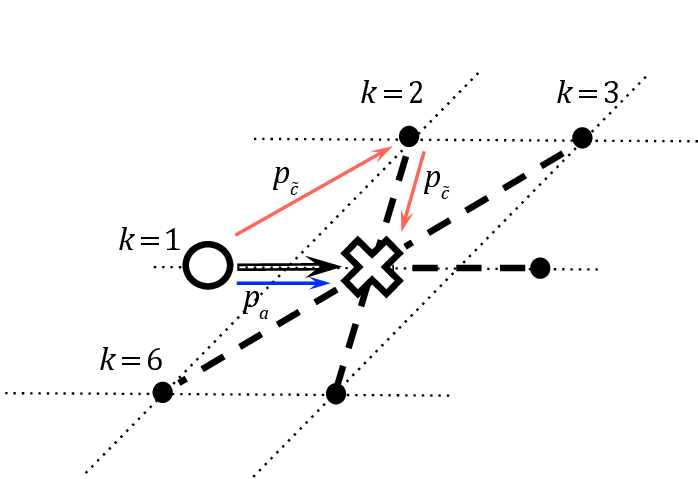}
  \caption{
    An example of the vacancy tracks for jump $l=a$.
    The tracer is positioned at the central site (open cross)
    just after moving as the double shafted arrow.
    When considering the vacancy tracks including
    one or two routes, the vacancy tracks are limited
    in the ones shown by red and blues allows.
  }
  \label{mae_a}
\end{figure}

Before solving eq.~\eqref{eq.aveCos} to obtain the average cosines
$\left\langle {\cos \theta } \right\rangle _{l = a,~\tilde c}^{\left( {n } \right)}$,
$\left\{ P^{(l)}_k\right\}$ should be prepared first, which represents
the probability that the tracer gets pulled back by the vacancy from site $k$,
after the tracer moved through route $l$.
$\left\{ P^{(l)}_k\right\}$ is given as the sum of the realization probabilities
for the vacancy tracks pulling back the tracer from site $k$.
This value can be reasonably approximated with considering only
the vacancy tracks having highest realization probabilities.
This is an example to obtain $\left\{ P^{(l)}_k\right\}$:
The tracer just moved to the central site (open cross) through
the route $a$ (double shafted arrow) in Fig.~\ref{mae_a}.
Suppose that the vacancy moves through a route $a$,~$\tilde c$
with the selection probability $p_{a,~\tilde c}$ (given later) and
$\left\{ P^{(l)}_k\right\}$ is given as the product of the ones
of the routes included in the vacancy track. Thus, the realization
probability becomes lower when the track consists of larger number
of the diffusion routes in general. Now, we consider the two vacancy
tracks shown as the red and blue arrows in Fig.~\ref{mae_a}.
The former consists of two routes $\tilde c$ and the latter does of
one route so the realization probabilities of the both tracks is
given as $p_{\tilde c}^2$ and $p_{a}$ respectively. Hence, $P^{(a)}_2$=$p_{\tilde c}^2+p_a$.

\begin{figure}
  \centering
  \includegraphics[width=\hsize]{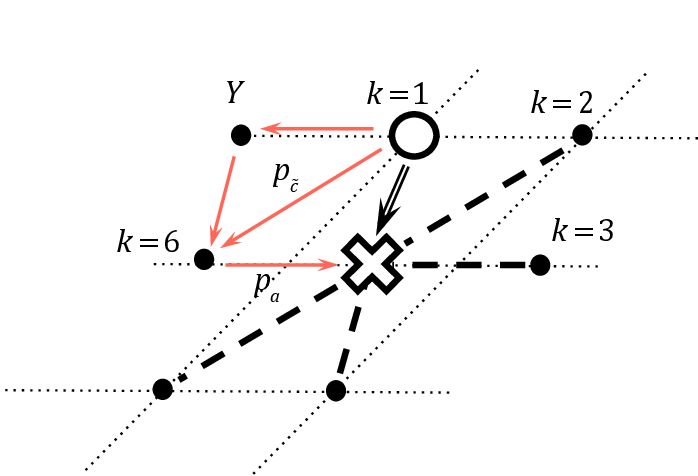}
  \caption{
    An example of the vacancy tracks for jump $l=tilde c$.
    The tracer is positioned at the central site (open cross)
    just after tracer moving as the double shafted arrow.
    The examples of the vacancy tracks are shown as the red arrows.
    The site $Y$ is newly defined to take into account
    the vacancy tracks including the path from $Y$ to 6.
  }
  \label{mae_c}
\end{figure}

\vspace{2mm}
$p_{a,~\tilde c}$ is in proportion to $\eta _{a,~\tilde c}(T)$ and given as 
\begin{eqnarray}
  p_{a,~\tilde c}(T) = \frac{{{\eta _{a,~\tilde c}(T)}}}{{2{\eta _a(T)} + 4{\eta _{\tilde c}(T)}}}.
  \label{eq.ac_select}
\end{eqnarray}
The denominator reflects that a rep-site connects to 2 rep-sites through
route $a$ and 4 rep-sites through route $\tilde c$. $p_{a,~\tilde c}$ depends on
temperature due to the Boltzmann factors in $\eta _{a,~\tilde c}$,
so the different vacancy tracks should be taken into account for different temperature.
We considered the high and low temperature cases separately.

\vspace{2mm}
First, for the high temperature case, the relationship,
$p_a \sim p_{\tilde c} \sim 1/6$~({\it e.g.}, $p_a/p_{\tilde c} = 1.03$ at $T$=1000~K),
is given from \textit{ab initio} calculations. Hence, $p_{l}=17\%$, $p_l^2 = 3.8\%$,
and $p_l^3 = 0.46$\% so the vacancy tracks including 3 moves or more would be ignored
Then, the vacancy cannot pulls back the tracer from site $k=3 \sim 5$ for the preceding
ion move through route $l=a$ and hence $P^{(a)}_{k=3\sim 5} = 0$.
The other $P^{(l=a)}_{k}$ are also straightforwardly given as $P^{(a)}_{k=1} = p_{a}$,
$P^{(a)}_{k=2} = p_{\tilde c}\cdot p_{\tilde c}$, and $P^{(a)}_{k=6} = p_{\tilde c}\cdot p_{\tilde c}$.
Applying the same analogy for $l=\tilde c$, $P^{(\tilde c)}_{k=3\sim 5}$ are zero
and the others are $P^{(\tilde c)}_{k=1} = p_{\tilde c}$, 
$P^{(\tilde c)}_{k=2} = p_{a} \cdot p_{\tilde c}$, and $P^{(\tilde c)}_{k=6} = p_{\tilde c}\cdot p_{a}$.

\vspace{2mm}
Secondly, for the low temperature case, $p_a$ is much greater than $p_{\tilde c}$
({\it e.g.}, $p_a/p_{\tilde c}~=~8.93$ at $T$=300K). Thus, we suppose that,
while the number of the vacancy moves through route $\tilde c$ is limited in 0 or 1,
that through route $a$ is not limited. Then, we have to consider the infinite patterns
of the vacancy tracks and it is apparently impossible. We introduced the following
three tools to count up the primary vacancy tracks which have the high realization probabilities:
\begin{itemize}
\item
  Suppose to consider the vacancy tracks as the vacancy being
  finally at the starting site after the arbitrary $n$ times of
  the vacancy moves through route $a$. The sum of the probabilities of
  the vacancy tracks is denoted by $\pi_{\rm{even}}$. 
\item
  Suppose to consider the vacancy tracks as the vacancy being
  finally at the left (right) site of the stating site after
  the arbitrary $n$ times of the vacancy moves.
  The sum of the probabilities of the vacancy tracks
  is denoted by $\pi_{\rm{odd}}$.
\item
  Suppose to consider the vacancy tracks as the vacancy being
  finally at the left (right) site of the stating site after
  the arbitrary 4n times of vacancy moves, imposing
  that the vacancy never gets into the sites positioned in the
  right (left) side from the starting site. The sum of the probabilities
  of the vacancy tracks is denoted as $\pi_{\rm{oneside}}$.
\end{itemize}
Here, $\pi_{\rm{even}}$ and $\pi_{\rm{oneside}}$ must be greater than 1
because the vacancy has to be the starting site for $n=0$.
It is explained in the appendix how to deduce $\pi_{\rm even}$,
$\pi_{\rm odd}$, and $\pi_{\rm oneside}$.
%

\vspace{2mm}
$P_k^{\left( l \right)}$ is deduced here for every ($k$,~$l$)
using the three tools:
{\bf (1)}~$l=a,~k=1$ case; We may consider the vacancy
tracks that the vacancy roams in the left side from site $k=1$
(including site $k=1$) through route $a$ before the pull-back.
The probability sum
\footnote{
  The sum of the probability for a group of vacancy tracks.
}
of the vacancy tracks before the pull-back
from site $k=1$ is exactly $\pi_{\rm{oneside}}$.
Hence, $P_{k = 1}^{\left( a \right)} = {\pi _{{\rm{oneside}}}}\cdot{p_a}$.
{\bf (2)}~$l=a,~k=2 \sim 6$ case; Two vacancy moves through route $\tilde c$
or more are required to pull back the tracer from these site.
Hence, $P_{k = 2 \sim 6}^{\left( a \right)}=0$.
{\bf (3)}~$l=\tilde c,~k=1$ case;
We may consider the vacancy tracks that the vacancy roams through
route $a$ before the pull-back.
The probability sum of the vacancy tracks before the pull-back
from site $k=1$ is exactly $\pi_{\rm{even}}$.
Hence, $P_{k = 1}^{\left( {\tilde c} \right)} = {\pi _{{\rm{even}}}}{p_c}$.
{\bf (4)}~$l=\tilde c,~k=2$ case; This case is quite similar to case (3).
The difference is just that the vacancy pulls back the tracer
from the site $k=2$ not $k=1$ after roaming through route $a$.
The probability sum just before the pull-back is exactly $\pi _{{\mathrm{odd}}}$.
Hence, $P_{k = 2}^{\left( {\tilde c} \right)} = {\pi _{{\mathrm{odd}}}}{p_c}$.
{\bf (5)}~$l=\tilde c,~k=3$ case;~We may only consider such vacancy tracks
that the vacancy moves to site $k=2$ from $k=3$.
The probability sum before the vacancy positioned at site $k=3$ is given
as $\pi_{\rm{odd}}p_c$ just like case (4), and that of the left vacancy tracks
is given as $\pi_{\rm{oneside}}p_a$ just like case (1).
$P_{k = 2}^{\left( {\tilde c} \right)}$ is eventually given as the product of them:
$P_{k = 2}^{\left( {\tilde c} \right)} = \pi_{\rm{odd}}p_c \cdot \pi_{\rm{oneside}}p_a$.
{\bf (6)}~$l=\tilde c,~k=3 \sim 5$ case;~
Any of the vacancy tracks have to include two moves through
route $\tilde c$ or more. Hence, $P_{k = 3 \sim 5}^{\left( \tilde c \right)}=0$.
{\bf (7)}~$l=\tilde c,~k=6$ case;~We may consider the two patterns
of vacancy tracks before the vacancy positioned at the site $k=6$:~
The vacancy moves to $k=1$ from $k=6$ or site $Y$ (shown in Fig.~\ref{mae_c}).
The probability sum before reaching site $k=6$ is given as
$\left( {{\pi _{{\rm{even}}}} + {\pi _{{\rm{odd}}}}} \right){p_{\tilde c}}$.
and that of the left vacancy tracks is equivalent to the one in case (1).
$P_{k = 6}^{\left( {\tilde c} \right)}$ is eventually given as the product of them:
$P_{k = 6}^{\left( {\tilde c} \right)} = \left( {{\pi _{{\rm{even}}}}
  + {\pi _{{\rm{odd}}}}} \right){p_{\tilde c}} \cdot {\pi _{{\rm{oneside}}}}{p_a}$.

\vspace{2mm}
We discussed the high temperature and low temperature cases separately above.
If considering the sum group of the vacancy tracks taken in the both cases
to calculate the probability sums $P_k^{\left( l \right)}$,
such $P_k^{\left( l \right)}$ may work well in the entire range of temperature.
The vacancy tracks for the high temperature case are actually included
in the ones of the low temperature case except for $l=\tilde a,~k=2,6$.
Eventually, $P_k^{\left( l \right)}$ is given as: 
\begin{eqnarray}
  P_k^{\left( l \right)} = \begin{cases}
    {\pi _{{\rm{oneside}}}} \cdot {p_a} & (l=a,~k=1)\\
    p_{\tilde c}^2 & (l=a,~k=2,6)\\
    {\pi _{{\rm{even}}}}\cdot{p_{\tilde c}} & (l=\tilde c,~k=1)\\
    {\pi _{{\rm{odd}}}} \cdot{p_{\tilde c}} & (l=\tilde c,~k=2)\\
    \pi_{\rm{odd}}\pi_{\rm{oneside}}\cdot p_a p_{\tilde c} & (l=\tilde c,~k=3)\\
    \left( {{\pi _{{\rm{even}}}} + {\pi _{{\rm{odd}}}}} \right){\pi _{{\rm{oneside}}}} \cdot {p_a}{p_{\tilde c}} & (l=\tilde c,~k=6) \\
    0 & (otherwise)
  \end{cases}
\end{eqnarray}
and eq.~\eqref{eq.aveCos} becomes
\begin{widetext}
\begin{eqnarray}
\left\langle {\cos \theta } \right\rangle _a^{\left( {n + 1} \right)}{\rm{ }} &=&  - P_{k = 1}^{\left( a \right)} \cdot \left\langle {\cos \theta } \right\rangle _a^{\left( n \right)} - 2 \cdot \frac{{{d_{\tilde c}}}}{{{d_a}}} \cdot P_{k = 2,6}^{\left( a \right)} \cdot \cos {\theta _{a\tilde c}} \cdot \left\langle {\cos \theta } \right\rangle _c^{\left( n \right)},\\
\left\langle {\cos \theta } \right\rangle _{\tilde c}^{\left( {n + 1} \right)} &=&  - \frac{{{d_a}}}{{{d_{\tilde c}}}} \cdot \left\{ {P_{k = 3}^{\left( {\tilde c} \right)} \cdot \cos \left( {2{\theta _{c\tilde c}} + {\theta _{a\tilde c}}} \right) + P_{k = 6}^{\left( c \right)} \cdot \cos {\theta _{a\tilde c}}} \right\} \cdot \left\langle {\cos \theta } \right\rangle _a^{\left( n \right)} - \left\{ {P_{k = 1}^{\left( {\tilde c} \right)} - P_{k = 2}^{\left( {\tilde c} \right)} \cdot \cos 2{\theta _{c\tilde c}}} \right\} \cdot \left\langle {\cos \theta } \right\rangle _{\tilde c}^{\left( n \right)},
\end{eqnarray}
\end{widetext}
from which the $n$th-order average cosigns $\left\langle {\cos \theta } \right\rangle _{a,~\tilde c}^{\left( n \right)}$
are given for any order $n$.

\section{Results and Discussion}
\label{res_dis}
\begin{figure}
  \centering
  \includegraphics[width=\hsize]{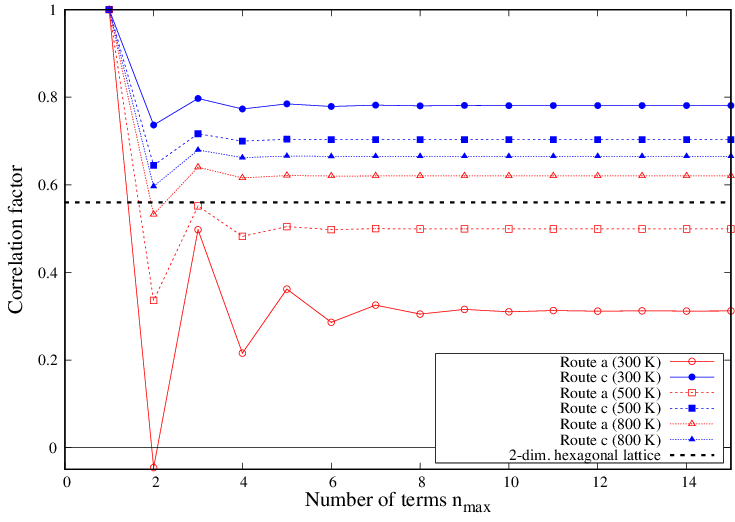}
  \caption{
    The convergence of correlation factor $f_{a,~\tilde c}$ for 300, 500, and 800~K
    evaluated from eq.~\eqref{eq.corrFac}.
    The black dotted line represents the correlation factor of
    2-dimensional hexagonal lattice and the value is 0.56.\cite{2007MEH}.
  }
  \label{fig.convergence}
\end{figure}

Fig.~\ref{fig.convergence} shows the convergence of $f_{a,~\tilde c}$ for $n_{\rm max}$.
$f_{a,~\tilde c}$ apparently vibrates for $n_{\rm max}$ because every pull-back denies
the preceding tracer move. Further on, we consider only the converged values,
which is actually used to solve the model.

\vspace{2mm}
If route $a$ and $\tilde c$ are equivalent,
the coarse grained structure shown in Fig.~\ref{fig.cu3ti_model}
is regarded as ideal 2-dimensional hexagonal lattice.
Its correlation factor is analytically known as 0.56\cite{2007MEH}
(shown as the dotted line in Fig.~\ref{fig.convergence}),
so it is interesting to compare $f_{a,\tilde c}$ with the ideal value.
First, in the case of low temperature ($T$=300~K),
it is observed that $f_a$ is lower and $f_{\tilde c}$
is higher than 0.56 respectively. This fact comes
from that the vacancy moves through routes $l=a$ much
more easily than through route $l=\tilde c$:
The vacancy easily goes away from the tracer through route $a$
after the tracer move $l=\tilde c$,
so the pull-back seldomly happens and $f_{\tilde c}$ is becomes higher.
On the other hand, for the tracer move $l=a$,
the vacancy moves through route $a$ on the same line for many times,
so the pull-back often happen and $f_{a}$ becomes lower.
In another way of thinking, the latter situation is similar to
the 1-dim diffusion so $f_{a}$ gets much lower.
Secondly, in the case of high temperature ($T$=1000~K),
it is expected that both $f_a$ and $f_{\tilde c}$ converge
around 0.56 when temperature increasing,
because route $a$ and $\tilde c$ are almost equivalent
at high temperature and the diffusion network is similar to
that on ideal 2-dim. hexagonal lattice.
The converged values are actually higher than 0.56 by $\sim$0.08
probably because of that the vacancy tracks including 3 routes and more are ignored,
which can cause the overesimation of the self-diffusion coefficient.
Nevertheless, the bias is quite small even at high temperature and presumably
it should get less at low temperature because the number of the route $a$
is not limited.

\begin{figure}
  \centering
  \includegraphics[width=\hsize]{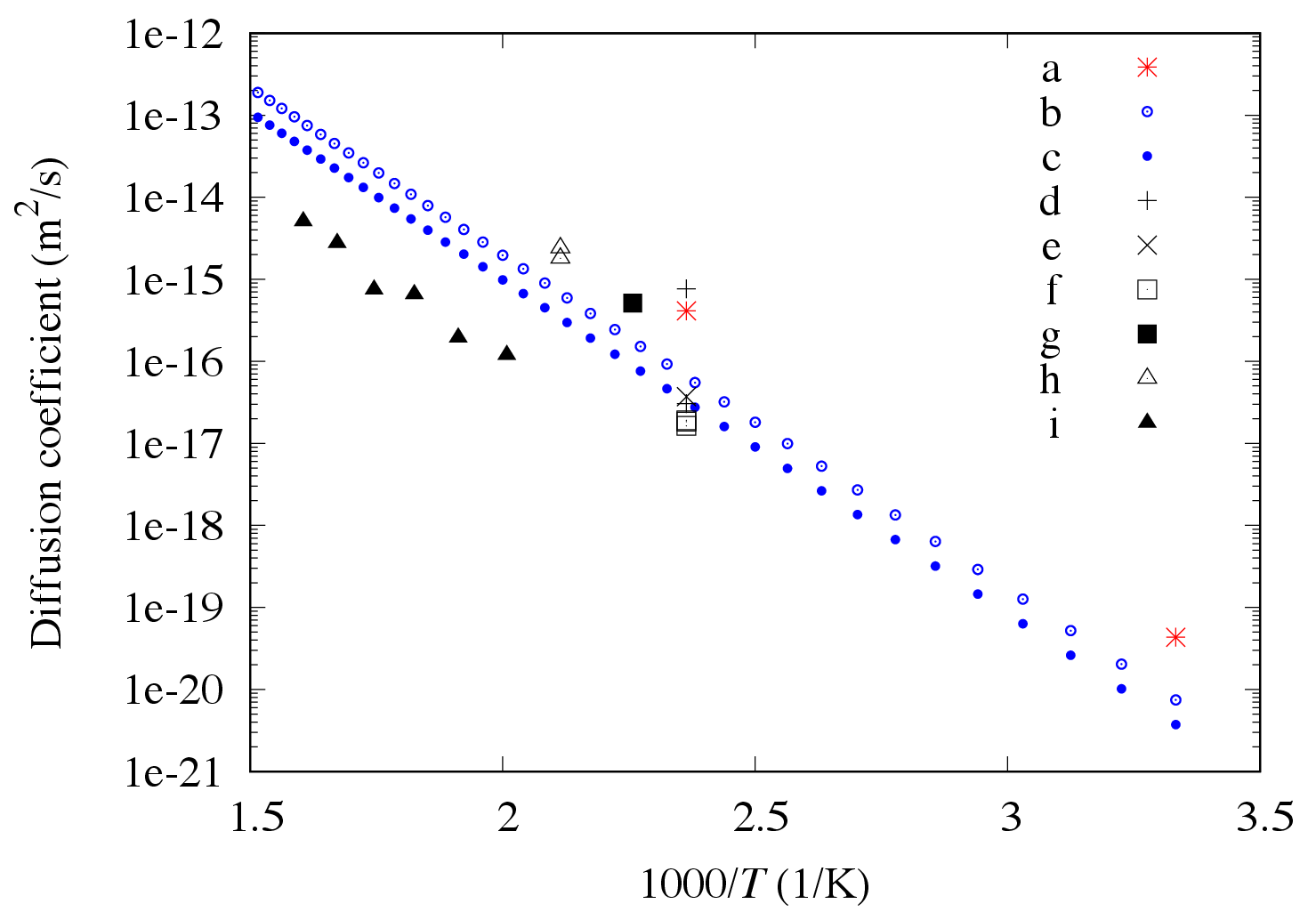}
  \caption{
    a: Classical MD \cite{2012GAO}, b: This work (M=$\inf$), c: This work (M=4),\newline
    d: Expt.\cite{2006CHA}, e: Expt.\cite{2009CHA}, f: Expt.\cite{2013YAN}, g: Expt.\cite{2013WAN},
    \newline h: Expt.\cite{2011KUM}, i: Expt.\cite{2011PAU}, \newline
    This graph compares our predictions for the diffusion coefficients (blue points)
    with the ones reported from the experiments (black points) and classical molecular
    dynamics (red points).
    The blue open (close) points correpond to the analysis with $M=\infty,~(4)$.
  }
  \label{fig.result}
\end{figure}

\vspace{2mm}
Once getting the correlation factors $f_{a,~\tilde c}$, ${D_\mathrm{C{u_3}Ti}}$
can be calculated from eq.~\eqref{cu3ti_diff}$\sim$\eqref{each_axis}.
Since $\varepsilon$-Cu$_3$Sn phase consists of Cu$_3$Ti-type and D0$_{19}$-type
domains with the ratio of $(M-2)$:2, the self-diffusion coefficient
in the phase is given as 
\begin{eqnarray}
  {D_{\varepsilon  - {\rm{C}}{{\rm{u}}_{\rm{3}}\rm{Sn}}}}
  = \frac{{M - 2}}{M}{D_{{\rm{C}}{{\rm{u}}_{\rm{3}}}{\rm{Ti}}}}\;\;
  \left(\because {D{0_{19}} = 0} \right).
\end{eqnarray}
The temperature dependency of $D_{\varepsilon  - {\rm{C}}{{\rm{u}}_{\rm{3}}\rm{Sn}}}$
is shown in Fig.~\ref{fig.result} for both $M=4$ and $M=\infty$.
Although they have twice different values, the difference is relatively
small compared with the scatter of the experimental values:~
Therefore, the experimental contradictions cannot be explained
from the difference of periodicity $M$.
It should be noted that the bias in the correlation factor $f$ by $\sim$0.08
is much smaller than the scatter.
Our results look reproducing the experimental values reasonably
as the predicted values pass though just the middle of the experimental values.
On the other hand, the previous classical MD \cite{2012GAO} overestimates
our results by a digit. It is possible that the MD underestimated the potential top
of diffusion route because of too large time step, since the time step is not shown
in the previous paper.\cite{2012GAO}
However, useally the time step is femto-second order and it is sufficiently small
compared with the time scale of the vibration shown in Tab.~\ref{tab.cu3ti}.
Hence, we tentatively conclude that the overestimation comes from the practical force-field,
which is presumely less reliable especialy for the description of the saddle point
since three ions or more interact with the tracer.

\section{summary and prospects}
\label{sec:conclusion}
We established a new modeling scheme for ion self-diffusion
coefficient by introducing some novel concepts and formalisms.
The scheme is applied to obtain the Cu self-idffusion coefficient
in $\varepsilon$-Cu$_3$Sn phase of Cu-Sn alloy.
The highlights are `domain division' and `coarse grainining'
of the diffusion network based on the barrier energies
predicted by the {\it ab initio} calculation.
In the former concept, the diffusion network is divided
into the disjunct domains. Then, the calculation for the original
large unitcell is replaced by the ones for the different types of
simple domains, which significantly reduces the calculation cost.
The diffusion network of our target system, $\varepsilon$-Cu$_3$Sn,
is divided into the two dypes of domains, which have 1- and
2-dimenstional networks respectively. It is concluded based on
the textbook\cite{2007MEH} that the former domain never
contributes to the self-diffusion. 
In the latter concept, the diffusion network of each domain
is coarse-grained by representing the ion sites with just a site,
which are connected by the routes having the lowest barrier energies.
This is on ground of that the vacancy moves through such routes much more
easily than through the others. Eventually, it is significantly
simplified to count up the vacnacy tracks.
After introducing the concepts, the model for $\varepsilon$-Cu$_3$Sn
phase was successfully established and much better preditions by a digit than the previous
classcial MD study\cite{2012GAO} in the comparison with the experimental values.

\vspace{2mm}
If the overesimation of self-diffusion coefficient
in the previous classical MD work\cite{2012GAO}
comes from the usage of the practical potential,
{\it ab initio} MD (AIMD) can be a strong competitor
of our modeling scheme. Nevetherless, our scheme is
superior to AIMD at the following points:
First, the quantities included in the model can be
evaluated from more reliable methods such as diffusion
Monte Carlo method.\cite{2001FOU}
The merit is remarkable especially for such systems
as transition metal dioxide\cite{2016LUO,2017TRA},
where the severe accuracy is required to evaluate
the electronic correlation effect.
Secondly, it is challenging to evaluate the self-diffusion
coefficient with AIMD in terms of the conputational cost,
because ion jump is a rare event.
The recent attempt using the color-diffusion algorithm
with non-equiblium MD realizes more efficient simulation
by increasing the rate of ion jumps.\cite{2016SAN}
Nevertheless, 1100 steps were used to obtain the self-diffusion
coefficient in body centered molybdenum crystal at 1600~K.
To make matters worse, more steps are required at lower temperature
because ion jump occurs less frequently. The application of AIMD
at low temperature is still challenging.
On the other hand, while our scheme requires to construct the model by hand,
the self-diffusion coefficient can be calculated with the reasonable cost.
The merit becomes remakable especially when performing a lot of
similar calculations: For example, surveying how the addicted ions
affect the self-diffusion coefficient with modeling the inclusion
of the addictinves using virtual crystal approximation.\cite{2000BEL}.

\section{Acknowledgments}
The computation in this work has been performed 
using the facilities of the Research Center for Advanced Computing Infrastructure (RCACI) at JAIST.
K.H. is grateful for financial support from a KAKENHI grant (JP17K17762), 
a Grant-in-Aid for Scientific Research on Innovative Areas ``Mixed Anion'' project (JP16H06439) from MEXT, 
PRESTO (JPMJPR16NA) and the Materials research by Information Integration Initiative (MI$^2$I) project 
of the Support Program for Starting Up Innovation Hub from Japan Science and Technology Agency (JST). 
R.M. is grateful for financial supports from MEXT-KAKENHI (17H05478 and 16KK0097), 
from Toyota Motor Corporation, from I-O DATA Foundation, 
and from the Air Force Office of Scientific Research (AFOSR-AOARD/FA2386-17-1-4049).
R.M. and K.H. are also grateful to financial supports from MEXT-FLAGSHIP2020 (hp170269, hp170220).


\section{Appendix}
\setcounter{equation}{0}
\renewcommand{\theequation}{\roman{equation}}
The probability sums  $\pi_{\rm{even}}$, $\pi_{\rm{odd}}$, and $\pi_{\rm{oneside}}$
are deduced in the appendix. First, $\pi_{\rm{even}}$ corresponds to
the probability sum of the following vacancy tracks:
After 2$n$ ($n=0,~1,~2,~\cdots$) moves, the vacanc is positioned
at the starting site.
The realization probability of one of them is given as 
$
{{p_a^{2n} \cdot \left( {2n} \right)!} \mathord{\left/
 {\vphantom {{p_a^{2n} \cdot \left( {2n} \right)!} {{{\left( {n!} \right)}^1}}}} \right.
 \kern-\nulldelimiterspace} {{{\left( {n!} \right)}^2}}}
$
and $\pi_{\rm{even}}$ is obtaied by summing up them:
\begin{eqnarray}
  {\pi _{{\rm{even}}}} = \sum\limits_{n = 0}^\infty  {{{p_a^{2n} \cdot \left( {2n} \right)!} \mathord{\left/
 {\vphantom {{p_a^{2n} \cdot \left( {2n} \right)!} {{{\left( {n!} \right)}^2}}}} \right.
 \kern-\nulldelimiterspace} {{{\left( {n!} \right)}^2}}}} .
\end{eqnarray}
Secondly, $\pi_{\rm{odd}}$ corresponds to the probability sum of the following
vacancy tracks: After 2$n$+1 ($n=0,~1,~2,~\cdots$) moves, the vacancy is
positined at the just right (left) site of the starting one.
The realization probability of one of them is given as
$
{{{p_a^{2n + 1} \cdot \left( {2n + 1} \right)!} \mathord{\left/
 {\vphantom {{p_a^{2n + 1} \cdot \left( {2n + 1} \right)!} {n!\left( {n + 1} \right)!}}} \right.
 \kern-\nulldelimiterspace} {n!\left( {n + 1} \right)!}}}
$
and $\pi_{\rm{odd}}$ is obtain by summing up them:
\begin{eqnarray}
  {\pi _{{\rm{odd}}}} = \sum\limits_{n = 0}^\infty  {{{p_a^{2n + 1} \cdot \left( {2n + 1} \right)!} \mathord{\left/
        {\vphantom {{p_a^{2n + 1} \cdot \left( {2n + 1} \right)!} {n!\left( {n + 1} \right)!}}} \right.
        \kern-\nulldelimiterspace} {n!\left( {n + 1} \right)!}}}.
\end{eqnarray}
Lastly, $\pi_{\rm{oneside}}$ corresponds to the probability sum
of the following vacancy track: After 2$n$+1 ($n=0,~1,~2,~\cdots$) moves,
the vacancy is positioned at the starting site with imposing that the vacancy
never goes to the right (left) side of the starting site during moving.
The total number of such tracks is given as the Catalan number\cite{1999MON}
\begin{eqnarray}
{c_n} = \frac{1}{{n + 1}}\frac{{\left( {2n} \right)!}}{{{{\left( {n!} \right)}^2}}}.
\end{eqnarray}
Thus, $\pi_{\rm{oneside}}$ is given as:
\begin{eqnarray}
  {\pi _{{\rm{oneside}}}} = \sum\limits_{n = 0}^\infty  {{c_n}p_a^n}.
\end{eqnarray}
\bibliographystyle{apsrev4-1}
\bibliography{references}
\end{document}